\documentclass[11pt]{article}
\usepackage{graphicx}
\usepackage{amsmath}
\usepackage{amsfonts}
\usepackage{amssymb}
\usepackage{MnSymbol}

\newcommand{\be}{\begin{equation}}
\newcommand{\ee}{\end{equation}}
\newcommand{\tr}{{\rm Tr}}
\newcommand{\Wg}{{\rm Wg}}

\begin{document}

\title{Statistics of time delay in quantum chaotic transport}
\author{Marcel Novaes\\
{\small Instituto de F\'isica, Universidade Federal de Uberl\^andia}\\ {\small Av. Jo\~ao
Naves de \'Avila 2121, Uberl\^andia, MG, 38408-100, Brazil}}\date{} \maketitle
\begin{abstract}
We study the statistical properties of the time delay matrix $Q$ in the context of
quantum transport through a chaotic cavity, in the absence of time-reversal invariance.
First, we approach the problem from the point of view of random matrix theory, and obtain
exact results that provide the average value of any polynomial function of $Q$. We then
consider the problem from the point of view of the semiclassical approximation, obtaining
the entire perturbation series for some energy-dependent correlation functions. Using these correlation
functions, we show agreement between the random matrix and the semiclassical approaches for several statistical properties.
\end{abstract}

\section{Introduction}

Quantum scattering processes at energy $E$ can be described by the scattering matrix
$S(E)$, which transforms incoming wavefunctions into outgoing wavefunctions. This matrix
is necessarily unitary, in order to enforce conservation of probability and,
consequently, conservation of charge. Another important operator is the Wigner-Smith time
delay matrix $Q$ \cite{wigner,smith}, a hermitian matrix related to the energy derivative of
$S$. Its eigenvalues are the delay times of the system, and its normalized trace is the Wigner time
delay, $\tau_W=\frac{1}{M}{\rm Tr}Q$. These quantities contain information about the time a particle spends
inside a scattering region. A thorough discussion can be found in the review
\cite{review2}.

We consider a scattering region (`cavity') inside of which the classical dynamics is
strongly chaotic, connected to the outside world by small, perfectly transparent, openings. This can be realized
in experiments with microwave cavities \cite{cavities1,cavities2,cavities3,cavities4},
quantum dots \cite{dots1,dots2,dots3,dots4} and compound nuclei \cite{nuclear}. In this
case, there is a well defined classical decay rate $\Gamma$, such that the total
probability of a particle to be found inside the cavity decays exponentially in time,
$\sim e^{-\Gamma t}$. The quantity $\tau_D=1/\Gamma$ is called the classical `dwell
time'.

In the semiclassical regime (when $\hbar\to 0$ and the electron wavelength is much
smaller than the cavity size), the $S$ and $Q$ matrices are strongly oscillating
functions of the energy and a statistical approach is advantageous. One such approach is
based on random matrix theory (RMT). Its main hypothesis is that $S$ behaves like a
random unitary matrix, distributed in the unitary group according to some probability
measure (in the presence of time-reversal invariance, $S$ must also be symmetric; we do
not consider that situation in this work). If the openings are perfectly transparent,
this distribution is the normalized Haar measure of the group. A typical ergodicity
hypothesis is that the energy average of an observable for a fixed system is equal to an
average over many different, yet similar, systems (ensemble average). We denote these two averages by the same symbol, $\langle \cdot\rangle$.
In particular, the average of the
Wigner time delay is equal to the classical dwell time \cite{caio1,caio2}, $\langle
\tau_W\rangle=\tau_D.$

The RMT approach has had much success in describing so-called transport statistics
\cite{beenakker,prb78mn2008,prb80bak2009,macedo}, such as conductance, shot-noise, their
variances, etc. RMT can be applied to time delay, but usually this is not done starting
from the $S$ matrix, but rather from the Hamiltonian of the system. This allows better
control of the energy dependence and calculation of correlation functions, but requires
mapping the problem to a nonlinear supersymmetric $\sigma$-model
\cite{sigma1,sigma2,sigma3,sigma4}.

On the other hand, Brouwer, Frahm and Beenakker \cite{brouwer} succeeded in finding the
joint probability distribution for the eigenvalues of the time delay matrix $Q$, let us
denote them by $\tau_1,...,\tau_M$, where $M$ is the total number of scattering channels.
This allowed the calculation of marginal distributions \cite{brouwer2}, distribution of
Wigner time delay  (for $M=2$ \cite{Mis2} and in the limit $M\gg 1$ \cite{majumdar}), and the ensemble average of
linear moments \cite{simm1,simm2}, \be \mathcal{M}_n=\frac{1}{M}{\rm
Tr}[Q^n]=\sum_{i=1}^M \tau_i^n.\ee A few more general, non-linear, moments have also been
computed \cite{simm3,garcia}. A recent review, also considering extension to non-ideal openings and other symmetry classes, can be found in \cite{dima}.

In the first part of this work, we advance the RMT approach to statistics of time delay,
obtaining an explicit formula for arbitrary moments of $Q$, i.e. quantities of the
kind \be\label{moments} \mathcal{M}_{n_1,n_2,...}=\frac{1}{M}{\rm Tr}[Q^{n_1}]\frac{1}{M}{\rm
Tr}[Q^{n_2}]\cdots,\ee for any finite set of positive integers $n_1,n_2,...$ This allows
the calculation of the average value of any observable which is polynomial in $Q$. Our
method starts from the result of \cite{brouwer} and is based on Schur function expansions
and determinant evaluations. Importantly, our results are not perturbative in the number
of channels, being valid at finite values of $M$.

A different way of treating the problem of quantum chaotic transport is the semiclassical
approximation, in which elements of the $S$ matrix are written as sums over classical
scattering trajectories \cite{jalabert}. Calculation of energy-averaged transport
statistics then require so-called action correlations, sets of trajectories having the
same total action, leading to constructive interference. Using only identical
trajectories and ergodicity arguments \cite{berry,cvitanovic,hannay} one can recover some semiclassical
large-$M$ asymptotics. Quantum corrections, important at finite $M$, can be related to
non-identical trajectories having close encounters \cite{sieber}, and may be obtained
systematically \cite{muller,greg,novaes1}.

The semiclassical approach has also been used to understand time delay. Interestingly, in
this case one can use the periodic orbits \cite{balian} that live in the fractal chaotic
saddle of the system \cite{altmann} (sometimes called `the repeller'). This approach was
used to compute correlation functions \cite{eckhardt,vallejos,KS1} up to first few orders
in perturbation theory in $1/M$. It is actually equivalent \cite{KS2} to the
one based on scattering trajectories \cite{caio3}. Berkolaiko and Kuipers treated the linear moments $\mathcal{M}_n$ semiclassically,
initially in the large-$M$ limit \cite{berko1} and later up to the first finite-$M$
corrections \cite{berko2}, showing agreement with the corresponding RMT predictions.
These works actually consider the more general problem of an energy-dependent correlation
function \be\label{correlation} C_n(\epsilon)=\frac{1}{M}{\rm Tr}\left[
S^\dag\left(E-\frac{\epsilon \hbar}{2\tau_D}\right)S\left(E+\frac{\epsilon
\hbar}{2\tau_D}\right)\right]^n,\ee from which the moments $\mathcal{M}_n$ can be
recovered by differentiation (yet another semiclassical approach to time delay, that avoids correlation
functions, has recently been introduced by Kuipers, Savin and Sieber \cite{new}).

In the second part of this work, we advance the semiclassical approach to the statistics of
time delay, deriving from it a formula for correlation functions $C_n(\epsilon)$. This
formula is a Taylor series in $\epsilon$, the coefficients of which are rational
functions of $M$ expressed as finite sums involving characters of the symmetric group and
Stirling numbers. Our method is an extension of a recently introduced semiclassical
matrix model for transport statistics \cite{novaes2}. The structure of our formula for $
C_n(\epsilon)$ suggests that the agreement between semiclassics and RMT holds exactly in
$M$ for all non-linear moments $\mathcal{M}_{n_1,n_2,...}$ (which we computed in the first
part). However, even though this can be checked in many cases using the computer, we come
short of explicitly showing it in full generality.

This paper is organized as follows. In the next Section we present and discuss our
results, before entering into details of calculations. Section 3 contains an exposition
of some preliminary material. Section 4 has the derivation of our random matrix theory
results for the general moments (\ref{moments}), while Section 5 contains our semiclassical
approach to the correlation function (\ref{correlation}).

\section{Results and Discussion}

We start by extending the RMT approach and computing all nonlinear statistics of the time
delay matrix. For example, the average value of the moments $\mathcal{M}_n$ were found in
\cite{simm1} for general number of channels $M$, but expressed as a sum with $M$ terms. Our results imply
the following simple general formula, which contains a sum with only $n$ terms:
\be\label{ourmoments} \langle\mathcal{M}_n\rangle=\tau_D^n\frac{M^{n-1}}{n!}\sum_{k=0}^{n-1}(-1)^k{n-1
\choose k}\frac{[M-k]^n}{[M+k]_n},\ee where \be [x]^n=x(x+1)\cdots(x+n-1), \quad
[x]_n=x(x-1)\cdots(x-n+1),\ee are the raising and falling factorials.

As another example, the first four cumulants of the Wigner time delay were computed in
\cite{simm3} using some nonlinear differential equation for their generating function.
This amounts to finding the value of $\langle \tau_W^j\rangle$ for $j$ up to 4. Our
results imply the explicit general formula \be \label{wigmom}\langle
\tau_W^n\rangle=\frac{\tau_D^n}{n!}\sum_{\lambda\vdash
n}d_\lambda^2\frac{[M]^{\lambda}}{[M]_{\lambda}},\ee where the sum is over all partitions
of $n$, the length of a partition $\lambda$ is denoted $\ell(\lambda)$ (these concepts
are discussed in Section 3) and \be\label{factorials}
[M]^\lambda=\prod_{i=1}^{\ell(\lambda)}[M-i+1]^{\lambda_i},\quad
[M]_\lambda=\prod_{i=1}^{\ell(\lambda)}[M+i-1]_{\lambda_i}\ee are generalizations of the
rising and falling factorials. The quantity $d_\lambda$ is the dimension of the
irreducible representation of the permutation group labeled by $\lambda$, and it is given
by \be\label{dimension}
d_\lambda=n!\prod_{i=1}^{\ell(\lambda)}\frac{1}{(\lambda_i-i+\ell(\lambda))!}\prod_{j=i+1}^{\ell(\lambda)}
(\lambda_i-\lambda_j-i+j).\ee

The above examples are derived from particular cases of our most general result, which is
the following

{\bf Theorem:} \emph{Let $Q$ be the $M$-dimensional time delay matrix of a chaotic cavity with no time-reversal symmetry. Let $\lambda\vdash n$ and let $s_\lambda(Q)$ be a Schur function of matrix argument. Then, in terms of the quantities defined above, we have} \be\label{RMTSchur}  \langle
s_\lambda(Q)\rangle=(M\tau_D)^n\frac{d_\lambda}{n!}\frac{[M]^{\lambda}}{[M]_{\lambda}}.\ee

The functions $s_\lambda(Q)$ are actually homogeneous symmetric polynomials in the eigenvalues of $Q$. Since any symmetric polynomial in these variables
can be expressed as a linear combination of Schur functions, this can be seen as a
complete solution to the problem of computing the average value of polynomial (or analytic functions, if we allow infinite series) of
$Q$, such as the quantities $\mathcal{M}_{n_1,n_2,...}$ defined in (\ref{moments}). For instance, the first of these which are neither of the form (\ref{ourmoments}) nor of the form (\ref{wigmom}) are \be\label{M21} \langle \mathcal{M}_{2,1}\rangle=\frac{2M^2(M^2+2)}{(M^2-1)(M^2-4)},\ee and \be\label{M22} \langle \mathcal{M}_{2,2}\rangle=\frac{4M^2(M^4+8M^2-3)}{(M^2-1)(M^2-4)(M^2-9)},\quad \langle \mathcal{M}_{3,1}\rangle=\frac{6M^2(M^2+1)^2}{(M^2-1)(M^2-4)(M^2-9)}.\ee

In the second part of this work, we develop a new formulation for the semiclassical
approach to time delay. Following our previous work on transport statistics
\cite{novaes2}, this is based on a matrix integral which is designed to have the correct
diagrammatic expansion. In this way, we find for example that \be\label{C1}
C_1=\frac{1}{1-i\epsilon}-\frac{\epsilon^2}{M^2(1-i\epsilon)^5}-
\frac{\epsilon^2(1+12i\epsilon-8\epsilon^2)}{M^4(1-i\epsilon)^9}+O(1/M^6),\ee
and \be\label{C2}
C_2=\frac{(1-2i\epsilon-2\epsilon^2)}{(1-i\epsilon)^4}-
\frac{\epsilon^2(4+8i\epsilon-7\epsilon^2-2i\epsilon^3)}{M^2(1-i\epsilon)^8}+O(1/M^4).\ee
The leading order part of these functions appear in the Appendix of \cite{berko1}.

Solving exactly our matrix integral, we arrive at our most general result, which is a formula for the
correlation functions in the form of a Taylor series: \be
C_n(\epsilon)=\frac{1}{Mn!}\sum_{m=0}^\infty \frac{(Mi\epsilon)^m}{m!}\sum_{\lambda\vdash
n}\sum_{\mu\vdash m}d_\lambda
d_\mu\chi_\lambda(n)\frac{[M]^\lambda}{[M]_\mu}F_{\lambda,\mu},\ee where $\chi$ are the
characters of the permutation group (see Section 3) and $F_{\lambda,\mu}$ is some
nontrivial function for which we have an explicit form (see Section 5.4.1).

Following \cite{berko1}, the average value of moments $\mathcal{M}_m$ can be obtained as
\be \langle \mathcal{M}_m\rangle=\frac{\tau_D^m}{i^mm!}\left[\frac{d^m}{d\epsilon^m}\sum_{n=1}^m
(-1)^{m-n}{m \choose n}C_n(\epsilon)\right]_{\epsilon=0}. \ee These quantities have been
computed semiclassically up to the first few orders in perturbation theory in $1/M$ in
\cite{berko2}. Using the above expression for $C_n(\epsilon)$ we could compute them in closed
form as rational functions of $M$ up to $m=8$ and check that the results agree with the
RMT prediction (\ref{ourmoments}). Unfortunately, we could not establish this agreement
in general, because of the complicated nature of the function $F_{\lambda,\mu}$.

As any reader who compares Sections 4 and 5 will notice, the semiclassical calculation is
much more complicated than the RMT one, so much so that it may seem hardly worth it. We
can raise two points in its defence. First, it provides the energy-dependent correlation
functions, which have more information than the energy-independent RMT statistics
(\ref{RMTSchur}). For instance, correlation functions are required in order to develop a
semiclassical treatment of Andreev systems, in line with \cite{andreev1,andreev2}.
Second, the semiclassical approximation is in principle able to go beyond RMT by
including Ehrenfest time effects (see e.g. \cite{ehren1,ehren2,ehren3,ehren4}). These
possible developments are outside the scope of the present work, but we hope they will
attract attention in the future.

A last remark about our semiclassical calculation. It is based on an integral over
$N$-dimensional complex matrices, and requires that we take the limit $N\to 0$. This limit is needed to
enforce that our semiclassical expansions do not contain periodic orbits. It is
easily taken in the perturbative framework (see Section 5.2), i.e. order by order in
$1/M$. However, we cannot rigorously justify it for the exact calculation. This is why
we do not claim our semiclassical results as theorems. We believe the nature of this
limit is an interesting open problem that deserves further study.

\section{Preliminaries}
\subsection{Partitions and permutations}

A weakly decreasing sequence of positive integers, $\lambda = (\lambda_1, \lambda_2,...)$
is called a partition of $n$, denoted by $\lambda\vdash n$ or by $|\lambda|=n$, if $\sum_i\lambda_i=n$. Each of
the integers is a part, and the total number of parts is the length $\ell(\lambda)$.

Partitions of $n$ label the conjugacy classes of the permutation group $S_n$: the cycle
type of a permutation $\pi$ is a partition whose parts are the lengths of the cycles of
$\pi$, and two permutations $\pi,\sigma$ have the same cycle type if and only if they are
conjugated, i.e. if there exists $\tau$ such that $\pi=\tau\sigma\tau^{-1}$. Let
$\mathcal{C}_\lambda$ denote the set of permutations with cycle type $\lambda$, and
$|\mathcal{C}_\lambda|$ the number of elements in $\mathcal{C}_\lambda$.

The number of permutations in $S_n$ which have exactly $k$ cycles is the (unsigned)
Stirling number of the first kind, $[\begin{smallmatrix}n \\ k \end{smallmatrix}]$. These
numbers also appear when we expand the rising factorial, \be\label{Stir}
[x]^n=\sum_{k=0}^n\begin{bmatrix}n \\ k \end{bmatrix}x^k.\ee

For any finite group, there are as many irreducible representations as there are
conjugacy classes. Therefore, partitions of $n$ also label the irreducible
representations of $S_n$. The trace of permutation $\pi$, in the representation labeled
by $\lambda$, is denoted as $\chi_\lambda(\pi)$ and called its character. The character
of the identity, $\chi_\lambda(1)=d_\lambda$, is the dimension of the representation, for
which there is the explicit formula (\ref{dimension}). Characters of $S_n$ are class
functions, i.e. $\chi_\lambda(\pi)$ depends only on the cycle type of $\pi$ and we may
write $\chi_\lambda(\mu)$ if $\pi\in\mathcal{C}_\mu$. Characters satisfy
orthogonality relations, \be \sum_{\tau\in S_n}\chi_\mu(\tau) \chi_\lambda(\tau\sigma)=
\frac{n!}{d_\lambda}\chi_\lambda(\sigma)\delta_{\mu,\lambda}.\ee

\subsection{Symmetric functions}

Let $X$ be a matrix of dimension $N$, with eigenvalues $x_i$, $1\leq i\leq N$. Power sum
symmetric functions of matrix argument are defined as \be
p_\lambda(X)=\prod_{i=1}^{\ell(\lambda)} p_{\lambda_i}(X),\quad  p_n(X)={\rm
Tr}[X^n]=\sum_{i=1}^Nx_i^n.\ee They are clearly symmetric functions of the eigenvalues.

Another important family of symmetric functions are Schur functions, related to power
sums by \be\label{power2schur} s_\lambda(X)=\frac{1}{n!}\sum_{\mu\vdash
n}|\mathcal{C}_\mu|\chi_\lambda(\mu)p_\mu(X), \quad p_\lambda(X)=\sum_{\mu\vdash
n}\chi_\mu(\lambda)s_\mu(X).\ee These functions can also be written as a ratio of
determinants, \be\label{schurasdet}
s_\lambda(X)=\frac{\det(x_i^{\lambda_j-j+N})}{\Delta(X)},\ee where \be
\Delta(X)=\det(x_i^{j-1})=\prod_{i=1}^N\prod_{j=i+1}^N(x_j-x_i),\ee is the Vandermonde
determinant. The value of the Schur function when all arguments are equal to $1$ is
\be\label{s1N} s_\lambda(1^N)=\frac{d_\lambda}{n!}[N]^\lambda,\ee where $[N]^\lambda$ is
the generalization of the rising factorial defined in (\ref{factorials}). Noticing that
in the formula for $d_\lambda$ there appears the Vandermonde for $x_i=\lambda_i-i$, it is
also possible to show that \be \label{special}
\Delta(\{\lambda_i-i\})=s_\lambda(1^N)\prod_{j=1}^{N-1}j!.\ee

Let $d\vec{x}=dx_1\cdots dx_N$. In view of the identity \be\label{andreief} \int
d\vec{x}\det(f_i(x_j))\det(g_i(x_j)) = N!\det\left( \int dx f_i(x)g_j(x)\right),\ee
easily proved using the Leibniz formula for the determinant, the representation
(\ref{schurasdet}) of Schur functions shall be useful for performing multidimensional
integrals involving these functions.

\subsection{Weingarten functions}

Given $j=(j_1,j_2,...j_n)$, $m=(m_1,m_2,...,m_n)$ and $\tau\in S_n$, define the function \be
\delta_\tau[j,m]=\prod_{k=1}^n\delta_{j_km_{\tau(k)}}.\ee Let
$\mathcal{U}(N)$ be the group of $N\times N$ unitary complex matrices $U$ and let $dU$ denote
 its normalized Haar measure. Then the so-called Weingarten function of this group is defined by
\be \int dU \prod_{k=1}^nU_{a_kb_k}U^\dag_{c_kd_k}=\sum_{\sigma,\tau\in
S_n}\Wg_N(\tau\sigma^{-1})\prod_{k=1}^n \delta_\sigma[ad]\delta_\tau[bc],\ee where
$U^\dag$ denotes the transpose conjugate of $U$. The character expansion of this function
is known \cite{samuel,esposti,collins}, \be \Wg_N(g)=\frac{1}{n!}\sum_{\substack{\lambda\vdash n\\\ell(\lambda)\leq N}}
\frac{d_\lambda}{[N]^\lambda}\chi_\lambda(g),\ee and the orthogonality of characters
implies the following identity: \be\label{convo} \sum_{\tau\in
S_n}\chi_\lambda(\tau\theta)\Wg^{U}_N(\tau\sigma^{-1})=\frac{\chi_\lambda(\theta\sigma^{-1})}
{[N]^\lambda}.\ee

\section{Random Matrix Theory approach}

We wish to compute the average value of a Schur function of the time delay matrix,
$s_\lambda(Q)$. This will be done using the following result obtained in \cite{brouwer}:
if $\gamma=Q^{-1}$, then the probability distribution of this matrix is \be
P(\gamma)=\frac{1}{\mathcal{Z}}|\Delta(\gamma)|^2\det(\gamma)^Me^{-M\tau_D{\rm Tr}\gamma}.\ee
where \be \mathcal{Z}=\int_0^\infty |\Delta(\gamma)|^2\det(\gamma)^Me^{-M\tau_D{\rm
Tr}\gamma}d\gamma\ee is a normalization constant.

Let $\tau_i$, $1\leq i\leq M$ be the eigenvalues of $Q$ and $\gamma_i=1/\tau_i$ be the
eigenvalues of $\gamma$. The normalization constant is computed using
Eq.(\ref{andreief}):\be\mathcal{Z}=\int_0^\infty
d\vec{\gamma}\det(\gamma_j^{M+i-1}e^{-M\tau_D\gamma_j})\det(\gamma_i^{j-1})=
\frac{M!}{(M\tau_D)^{2M^2}}\det((M+j+i-2)!).\ee Standard determinant manipulations yield \be
\mathcal{Z}=\frac{1}{(M\tau_D)^{2M^2}}\prod_{j=1}^M j!(M+j-1)!.\ee

The quantity we are after is \be \langle
s_\lambda(Q)\rangle=\frac{1}{\mathcal{Z}}\int_0^\infty d\vec{\gamma}
|\Delta(\gamma)|^2\det(\gamma)^Me^{-M\tau_D{\rm Tr}\gamma}s_\lambda(\gamma^{-1}).\ee
Writing the Schur function as a determinant, as in Eq.(\ref{schurasdet}), and using the
following identity for the Vandermonde, \be
\Delta\left(\gamma^{-1}\right)=\frac{(-1)^{M(M-1)/2}\Delta(\gamma)}{\det\gamma^{M-1}},\ee
we arrive at \be \langle
s_\lambda(Q)\rangle=\frac{(-1)^{M(M-1)/2}}{\mathcal{Z}}\int_0^\infty d\vec{\gamma}
\det(\gamma_j^{2M+i-2}e^{-M\tau_D\gamma_j})\det(\gamma_i^{-\lambda_j+j-M}).\ee
Using Eq.(\ref{andreief}) again we have \be\label{intermediate} \langle
s_\lambda(Q)\rangle=\frac{(-1)^{M(M-1)/2}}{\mathcal{Z}(M\tau_D)^{2M^2-n}}M!\det((M-\lambda_j+j+i-2)!).\ee

Consider the determinant $\det((x_j+i)!)$. Suppose we factor out a term $(x_j+1)!$ from
each row. The remaining determinant has the following structure: its $ij$ element is a
monic polynomial in $x_j$ of degree $i-1$. It is well known that it therefore must be
equal to the Vandermonde $\Delta(x)$. Applying this argument to (\ref{intermediate}) we
get \be \langle
s_\lambda(Q)\rangle=\frac{1}{\mathcal{Z}(M\tau_D)^{2M^2-n}}s_\lambda(1^M)\prod_{j=1}^Mj!(M-\lambda_j+j-1)!,\ee
where we used $ \Delta(\{M-\lambda_i+i-2\})=(-1)^{M(M-1)/2}\Delta(\{\lambda_i\})$ and the
special value of the Vandermonde, Eq. (\ref{special}). Plugging in the values of
$s_\lambda(1^M)$ and $\mathcal{Z}$, we get \be \langle
s_\lambda(Q)\rangle=(M\tau_D)^n\frac{d_\lambda}{n!}[M]^\lambda\prod_{j=1}^M
\frac{(M-\lambda_j+j-1)!}{(M+j-1)!},\ee or, in terms of the generalized falling factorial
defined in (\ref{factorials}), our claimed result, \be\langle
s_\lambda(Q)\rangle=(M\tau_D)^n\frac{d_\lambda}{n!}\frac{[M]^\lambda}{[M]_\lambda}.\ee

The relation between power sums and Schur functions, Eq. (\ref{power2schur}), allows the
calculation of more familiar quantities, such as \be\label{Mn} \langle
\mathcal{M}_n\rangle=\frac{1}{M}\langle p_{n}(Q)\rangle=\frac{1}{M}\sum_{\lambda\vdash n}\chi_\lambda(n)\langle
s_\lambda(Q)\rangle.\ee The character $\chi_\lambda(n)$ is different from zero only if
$\lambda=(n-k,1^k)$ (so-called hook partitions), and is equal to $(-1)^k$ in this case.
On the other hand, the dimension $d_\lambda$ becomes ${n-1 \choose k}$ for hooks, and
with this we arrive at our example (\ref{ourmoments}). The other example we mentioned in Section 2 was
\be \langle \tau_W^n\rangle=\frac{1}{M^n}\langle p_{(1,1,...,1)}(Q)\rangle=\frac{1}{M^n}\sum_{\lambda\vdash
n}d_\lambda\langle s_\lambda(Q)\rangle.\ee Finally, consider the general moments $\mathcal{M}_{n_1,n_2,...}$. Without any loss of generality, we may assume that $\mu=(n_1,n_2,...)$ is a partition of some integer, $|\mu|$. Then, we have \be \langle \mathcal{M}_{n_1,n_2,...}\rangle=\frac{1}{M^{|\mu|}}\sum_{\lambda\vdash|\mu|}\chi_\lambda(\mu)\langle s_\lambda(Q)\rangle.\ee Using this expression, we recover our examples (\ref{M21}) and (\ref{M22}).

\section{Semiclassical approach}

In the semiclassical limit $\hbar\to 0$, $M\to \infty$, the element $S_{oi}$ of the $S$
matrix may be approximated by a sum over trajectories $\gamma$ starting at channel $i$
and ending at channel $o$ \cite{jalabert}: \be S_{oi}=\frac{1}{\sqrt{T_H}}\sum_{\gamma:i\to o}A_\gamma
e^{i\mathcal{S}_\gamma/\hbar}.\ee The phase $\mathcal{S}_\gamma$ is the action of
$\gamma$, while $A_\gamma$ is related to its stability. The prefactor contains the
so-called Heisenberg time, $T_H=M\tau_D$.

Consider the correlation function $C_n(\epsilon)=\frac{1}{M}{\rm Tr}\left[
S^\dag\left(E-\frac{\epsilon \hbar}{2\tau_D}\right)S\left(E+\frac{\epsilon
\hbar}{2\tau_D}\right)\right]^n$. Expanding the trace, we find a multiple sum over
trajectories, \be\label{multiple}
C_n=\frac{1}{MT_H^n}\prod_{k=1}^n\sum_{i_k,o_k}\sum_{\gamma_k,\sigma_k}A_\gamma A^*_\sigma
e^{i(\mathcal{S}_\gamma-\mathcal{S}_\sigma)/\hbar}e^{\frac{i\epsilon}{2\tau_D}(T_\gamma+T_\sigma)},\ee
such that $\gamma_k$ goes from $i_k$ to $o_k$, while $\sigma_k$ goes from $i_k$ to
$o_{k+1}$, i.e. $\sigma$ trajectories implement a cyclic permutation on the labels of the
channels. The channels labels are all being summed from $1$ to $M$.

In (\ref{multiple}) we have used \be \mathcal{S}_\gamma(E+\frac{\epsilon
\hbar}{2\tau_D})\approx \mathcal{S}_\gamma(E)+\frac{\epsilon \hbar}{2\tau_D}T_\gamma,\ee
where $T_\gamma$ is the total duration of $\gamma$. The quantity $A_\gamma=\prod_k
A_{\gamma_k}$ is a collective stability, while $\mathcal{S}_\gamma=\sum_k
\mathcal{S}_{\gamma_k}$ and $T_\gamma=\sum_k T_{\gamma_k}$ are the collective action and
duration of the $\gamma$ trajectories, and analogously for $\sigma$.

The result of the sum (\ref{multiple}) is, for a chaotic system, a strongly fluctuating
function of the energy. A local energy average is thus introduced which, under the
stationary phase approximation, requires $\gamma$ and $\sigma$ to have almost the same
collective action. In the past years \cite{sieber}, it has been established that these
action correlations arise when each $\sigma$ follows closely a certain $\gamma$ for a
period of time, and some of them exchange partners at so-called encounters. A
$q$-encounter is a region where $q$ pieces of trajectories run nearly parallel and $q$
partners are exchanged. This theory has been presented in detail in \cite{muller,R2c}. We
consider only systems not invariant under time-reversal, so $\sigma$ trajectories never
run in the opposite sense with respect to $\gamma$ trajectories.

\begin{figure}
\includegraphics[scale=0.7,clip]{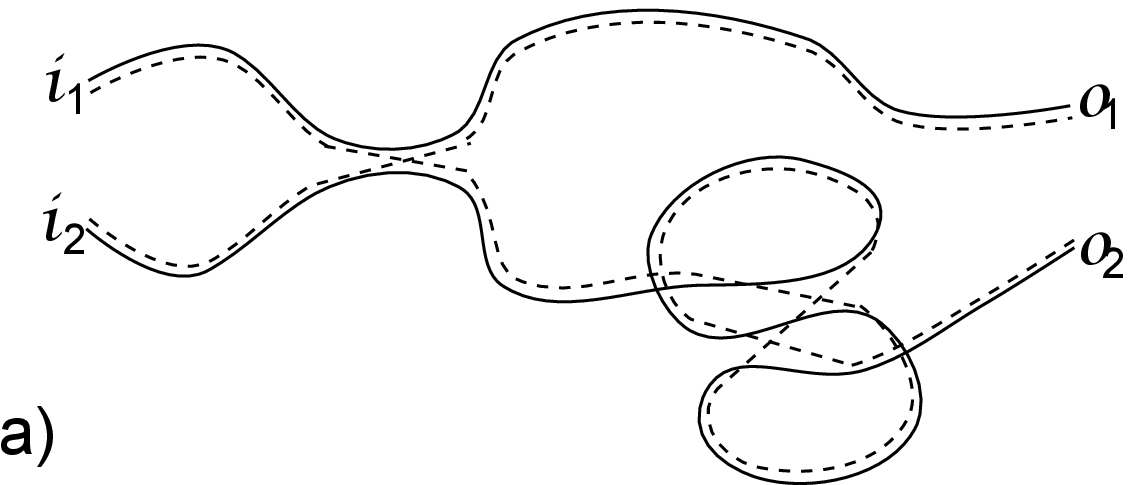}\qquad\includegraphics[scale=0.7,clip]{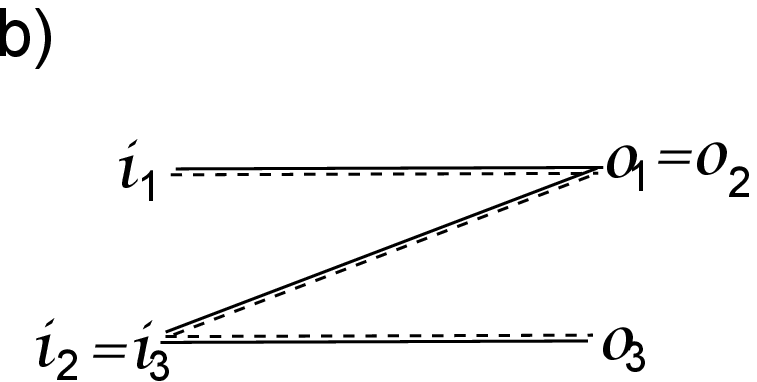}
\caption{a) Correlated trajectories contributing to $C_2$. Solid lines are $\gamma_1$ (going
from $i_1$ to $o_1$) and $\gamma_2$ (going from $i_2$ to $o_2$), dashed lines are
$\sigma_1$ (going from $i_1$ to $o_2$) and $\sigma_2$ (going from $i_2$ to $o_1$). In
this situation we have one $2$-encounter and one $3$-encounter (the encounters are greatly
magnified). b) Correlated trajectories contributing to $C_3$, in a case with coinciding channels. In both figures the chaotic nature of the trajectories is not shown.}
\end{figure}

For example, we show in Figure 1a a situation contributing to the second correlation
function, $C_2(\epsilon)$. Trajectory $\gamma_1$ starts in channel $i_1$ and ends in
channel $o_1$, while $\gamma_2$ starts in channel $i_2$ and ends in channel $o_2$. On the
other hand, $\sigma_1$ and $\sigma_2$ are initially almost identical to $\gamma_1$ and
$\gamma_2$, respectively, but they exchange partners in a $2$-encounter. Later,
$\gamma_2$ has a $3$-encounter with itself, inside of which the pieces of $\sigma_1$ are
connected differently. We also show in Figure 1b a situation contributing to $C_3(\epsilon)$ which has no encounters, but has coinciding channels. There are two major simplifications done here for visual clarity: 1) The encounters are greatly magnified, to show their internal structure; 2) The actual trajectories are extremely convoluted and chaotic. Many other examples of correlated trajectories can be found in previous work such as \cite{sieber,muller,greg,novaes1,berko1,berko2,novaes2}.

Correlated sets of trajectories contributing to the semiclassical calculation of
correlation functions can be depicted in the form of ribbon graphs, as suggested in
\cite{GregJack1,GregJack2}. The $q$-encounters become vertices of valence $2q$. Channels
also become vertices, but their valence depends on whether there are coinciding channels or not. The pieces of trajectories connecting vertices become
fat edges, or ribbons. Each ribbon is bordered by one $\gamma$ and one $\sigma$, and
these trajectories traverse the encounter vertices in a well defined rotation sense: a trajectory arriving from one ribbon departs via the adjacent ribbon (graphs endowed with a cyclic order around vertices are also called maps). We show in
Figure 2 the ribbon graphs corresponding to the trajectories shown in Figure 1.

\begin{figure}
\includegraphics[scale=0.75,clip]{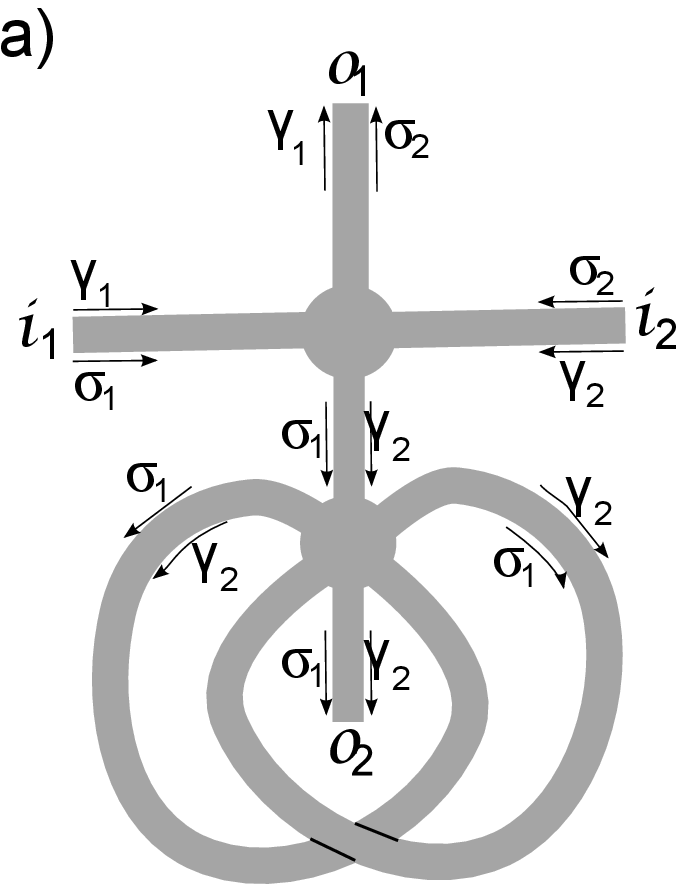}\qquad\includegraphics[scale=0.75,clip]{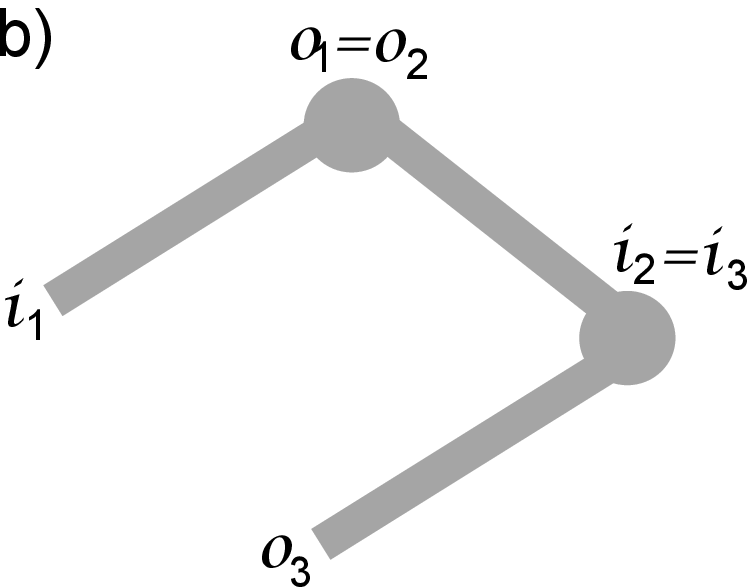}
\caption{The ribbon graphs corresponding to Figure 1. Each ribbon is bordered by one
$\gamma$ and one $\sigma$. Ribbons only meet at vertices, and $q$-encounters become
vertices of valence $2q$.}
\end{figure}

Following previous work on transport and on closed systems, Kuipers and Sieber obtained
some diagrammatic rules \cite{KS2}, that determine how much a given graph contributes to
the correlation function. The contribution of a graph factorizes into the contributions
of individual vertices and edges: an encounter vertex of valence $2q$ gives rise to
$-M(1-iq\epsilon)$; channels of any valence give rise to $M$; each ribbon gives rise to
$[M(1-i\epsilon)]^{-1}$. These rules were then used in several works dealing with time
delay statistics \cite{berko1,berko2,andreev1,andreev2}.

Notice that there are no periodic orbits in a ribbon graph that arises from the
semiclassical expansion of time delay. This means that we may start from $i_n$ and
follow $\sigma_1$ up to $o_1$, then follow $\gamma_1$ in reverse back to $i_1$, then
$\sigma_2$ to $o_2$, then $\gamma_2$ in reverse back to $i_2$, and so on, and
traverse every border of every ribbon exactly once. This means that the graph has a
single face.

The contribution of a graph will be proportional to $M^{V-E-1}$, where $V$ is the total
number of vertices (including channels) and $E$ is the total number of edges. The Euler
characteristic of a ribbon graph is $V-E+F$, where $F$ is the number of faces ($F=1$ in
our case). The Euler characteristic is also equal to $2-2g$, where $g$ is called the
genus. Therefore, the $1/M$ expansion coming from semiclassical diagrammatics is actually
what is called a genus expansion: the contribution of a graph is proportional to
$1/M^{2g}$. Graphs with $g=0$ are called planar (they can be drawn on the plane so that
the ribbons never cross each other), and they give the leading order contribution.

The graph in Figure 2a, for example, contributes \be
\frac{(1-2i\epsilon)(1-3i\epsilon)}{M^2(1-i\epsilon)^7}\ee to $C_2$. Notice that it is not
a planar graph, since there is a crossing between two of the ribbons. This particular
graph actually has $g=1$ (this means it may be drawn on a torus without any crossings). The graph in Figure 2b, on the other hand, is planar and contributes $(1-i\epsilon)^{-3}$ to $C_3$.

\subsection{Gaussian integrals and Wick diagrammatics}

We shall introduce a certain Gaussian matrix integral and formulate it diagrammatically,
using Wick's rule. This procedure has been discussed in detail for hermitian matrices in
\cite{difrancesco} and in \cite{bouttier}. The only difference compared to the present
work is that we integrate over non-hermitian matrices. Our diagrams are then interpreted
as providing the semiclassical formulation of the time delay problem. The same approach
was used to treat transport statistics in \cite{novaes2}.

Let $Z$ denote a general complex matrix of dimension $N$, and define \be\label{ave}
\llangle f(Z,Z^\dag)\rrangle\equiv \frac{1}{\mathcal{Z}}\int dZ
e^{-\Omega\tr[ZZ^\dag]}f(Z,Z^\dag), \ee where the normalization constant (not to be
confused with the normalization constant of Section 4) is \be \mathcal{Z}=\int dZ
e^{-\Omega\tr[(ZZ^\dag)]}.\ee We see (\ref{ave}) as an average value, but we use the
symbol $\llangle\cdot\rrangle$ to differentiate it from the true physical average we
considered in previous sections. For example, since the elements are actually
independent, it is clear that \be\label{cov} \llangle
Z_{mj}Z^\dag_{qr}\rrangle=\frac{\delta_{mr}\delta_{jq}}{\Omega}.\ee

Integrals over a product of matrix elements can be computed using the so-called Wick's
rule, which states that we must sum, over all possible pairings between $Z$'s and
$Z^\dag$'s, the product of the average values of the pairs. Namely,
\be\label{wick1}\left\llangle\prod_{k=1}^n
Z_{m_kj_k}Z^\dag_{q_{k}r_k}\right\rrangle=\sum_{\sigma\in S_n}\prod_{k=1}^n\llangle
Z_{m_kj_k}Z^\dag_{q_{\sigma(k)}r_{\sigma(k)}}\rrangle.\ee If we the quantity we wish to
average involves traces of $ZZ^\dag$, all we need to do is expand these traces in terms
of matrix elements and apply Wick's rule. Most importantly, we can then employ a
diagrammatic technique.

For example, suppose we wish to compute \be\label{avewick} \left\llangle {\rm
Tr}[(ZZ^\dag)^2]{\rm Tr}[(ZZ^\dag)^3]
Z_{i_1o_1}Z^\dag_{o_2,i_1}Z_{i_2o_2}Z^\dag_{o_1,i_2}\right\rrangle.\ee We start by
writing it as \be \sum_{m_1,...,m_5}\sum_{j_1,...,j_5}\left\llangle \left\{\prod_{k=1}^2
Z_{m_kj_k}Z^\dag_{j_k,m_{k+1}}\prod_{s=3}^5
Z_{m_sj_s}Z^\dag_{j_s,m_{s+1}}\right\}Z_{i_1o_1}Z^\dag_{o_2,i_1}Z_{i_2o_2}Z^\dag_{o_1,i_2}\right\rrangle,\ee
where all sums run from $1$ to $N$ (in the first product we mean $m_3\equiv m_1$, while
in the second product we mean $m_6\equiv m_3$). The diagrammatics consists in picturing
the matrix elements as pairs of arrows. Arrows that represent elements from $Z$ have a
marked end at the head, while arrows that represent elements from $Z^\dag$ have a marked
end at the tail. Arrows representing matrix elements coming from traces are arranged in
clockwise order around vertices, so that all marked ends are on the outside. Finally, the
elements that do not come from traces are arranged surrounding the other ones, also in
clockwise order. Since this is most easily explained by means of an image, we show it in
Figure 3(a).

Once we have arranged the arrows, Wick's rule consists in making all possible connections
between them, using the marked ends. Clearly, this produces a ribbon graph. According to
Eq.(\ref{cov}), when computing the value of a graph, each ribbon gives rise to a factor
$\Omega^{-1}$. For the example in Figure 3(a), there are 7! possible connections. We show
two of them in Figures 3(b,c). The coupling in Figure 3(b) leads to the identifications
\be i_1=m_1,\quad i_2=m_2=m_3=m_4=m_5 \quad o_1=j_2=j_3=j_4=j_5,\quad o_2=j_1,\ee and
gives a contribution of $\Omega^{-7}$ to the average (\ref{avewick}). Notice how this
coupling is similar to Figure 2. On the other hand, the coupling in Figure 3(c) leads to
the identifications \be i_1=m_1,\quad i_2=m_2=m_4=m_5 \quad o_1=j_2=j_3=j_5,\quad
o_2=j_1.\ee In this case the indices $m_3$ and $j_4$ remain free to be summed over.
Therefore, this coupling gives a contribution of $N^2\Omega^{-7}$ to the average
(\ref{avewick}).

\begin{figure}
\includegraphics[scale=0.7,clip]{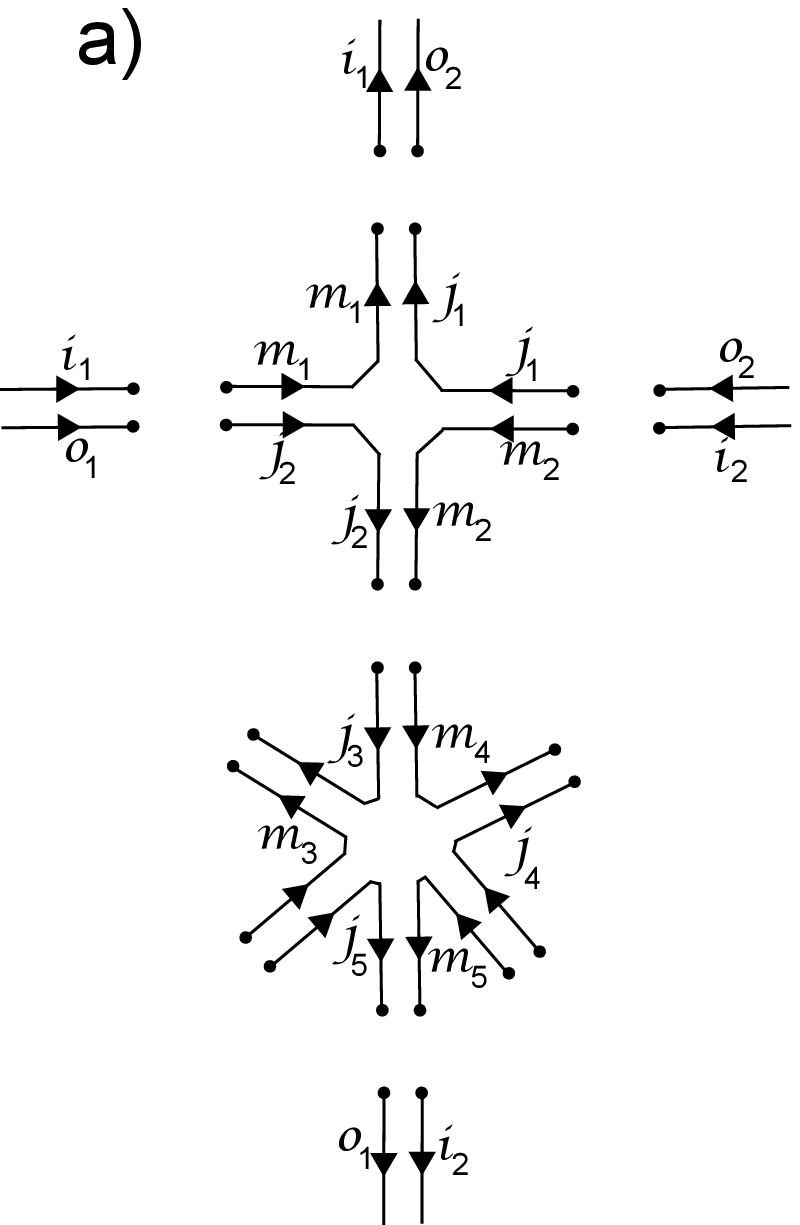}\quad \qquad\includegraphics[scale=0.55,clip]{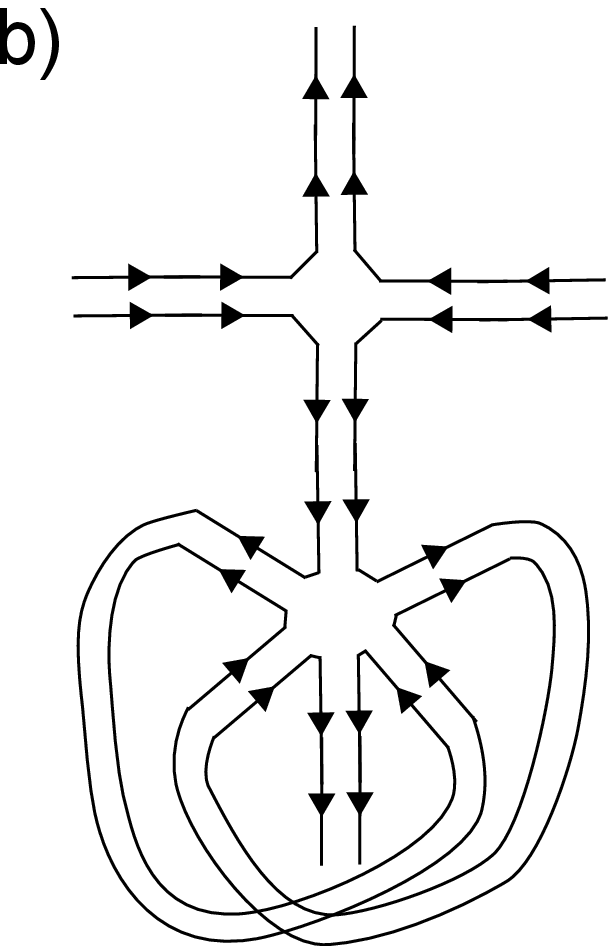}\quad \qquad\includegraphics[scale=0.55,clip]{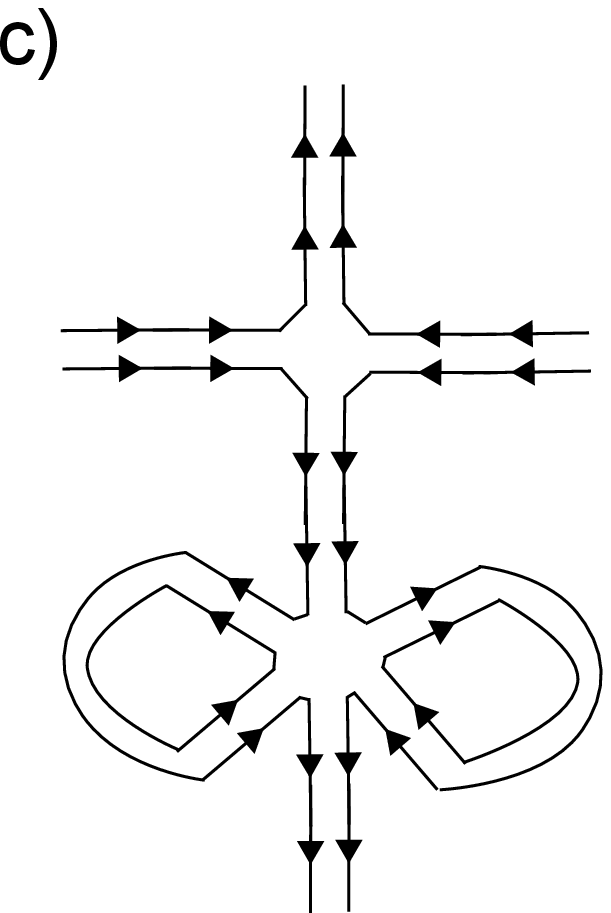}
\caption{Diagrammatics of Wick's rule, for the average in (\ref{avewick}). In a) we see
how the matrix elements are turned into arrows with marked ends and, in the case of traces, arranged
clockwise around vertices. In the vertex of valence 6 we have written each label only
once, for clarity. In b) and c) we see two particular Wick couplings, out of the possible
7!. The labels of the arrows in b) and c) are the same as in a). Notice the similarity
between b) and Figure 2.}
\end{figure}

Free indices arise from closed loops in the ribbon graph. Each such loop increases by one
the number of faces of the graph (every graph has at least one face). Therefore, the
power of $N$ in the contribution of a given coupling is always one less than the number
of faces in the graph.

It should be clear that this theory is very close to the semiclassical approach to time
delay, provided we choose $\Omega=M(1-i\epsilon)$. However, the ribbon graphs in the
semiclassical theory always have a single face. As we have just mentioned, this
corresponds to keeping only those Wick couplings whose contribution does not depend on
$N$. Since all contributions are proportional to a positive power of $N$, we could simply
let $N\to 0$.

\subsection{Matrix integrals for correlation functions}

Let $\xi=(1\,2\,\cdots \, n)$ be the cyclic permutation of the first $n$ positive integers,  and
let $\vec{i}=(i_1,...,i_n)$ and $\vec{o}=(o_1,...,o_n)$. Introduce the integral
\be\label{ourintegral} G_n(M,\epsilon,N,\vec{i},\vec{o})=\frac{1}{M\mathcal{Z}}\int dZ
e^{-M\sum_{q\ge 1} \frac{(1-iq\epsilon)}{q}\tr[(ZZ^\dag)^q]} \prod_{k=1}^n
Z_{i_ko_k}Z^\dag_{o_{\xi(k)}i_k}.\ee This can be seen as a Gaussian average as the ones
considered previously, if we understand the first term in the exponent,
$e^{-M(1-i\epsilon) \tr(ZZ^\dag)}$, to be part of the measure. Accordingly, we set \be\label{norm}
\mathcal{Z}=\int dZ e^{-M(1-i\epsilon)\tr[(ZZ^\dag)]}.\ee

The rest of the exponential can be Taylor expanded as \be e^{-M\sum_{q\ge 2}
\frac{(1-iq\epsilon)}{q}\tr[(ZZ^\dag)^q]}=\sum_{t=0}^\infty
\frac{(-M)^t}{t!}\left(\sum_{q\ge 2}
\frac{(1-iq\epsilon)}{q}\tr[(ZZ^\dag)^q]\right)^t.\ee For now, we consider this as a
formal power series and integrate term by term, employing Wick's rule and its
diagrammatical representation previously discussed. By construction, encounter vertices of valence $2q$ will be accompanied by the factor $-M(1-iq\epsilon)$, giving the correct
semiclassical diagrammatic rules.

The integral (\ref{ourintegral}) is therefore designed to automatically produce all the
required ribbon graphs for the semiclassical evaluation of the correlation function
$C_n$. The exponential produces all possible encounters, while the matrix elements in the
last product play the role of the channels. In line with Eq.(\ref{multiple}), we must sum
over all channels from $1$ to $M$, i.e. we must consider the quantity \be\label{sumio}
\mathcal{G}_n(M,N,\epsilon)=\sum_{\vec{i},\vec{o}}G_n(M,N,\epsilon,\vec{i},\vec{o})\equiv
\sum_{i_1,\cdots,i_n=1}^{M}\sum_{o_1,\cdots,o_n=1}^{M}G_n(M,N,\epsilon,\vec{i},\vec{o}).\ee

The matrix integral produces more graphs than needed, but we have provided for this
overcounting. For example, the Taylor series of the exponential naturally has a $t!$ in
the denominator, which is responsible for eliminating the symmetry associated with
shuffling the vertices, when there are $t$ of them.  Also, graphs are produced that
differ from each other only by the rotation of a vertex. This is why we have divided
$\tr[(ZZ^\dag)^q]$ by $q$: it remedies the overcounting that would be caused by the
possible $q$ rotations of the vertex.

As we have discussed, in order to select only those ribbon graphs with a single face it
is necessary to take the limit $N\to 0$ at the end of the calculation. Therefore, the
correlation function will be given by \be C_n(M,\epsilon)=\lim_{N\to
0}\mathcal{G}_n(M,N,\epsilon).\ee

It is not very difficult to implement Eq.(\ref{ourintegral}) in a computer and obtain the
first few orders in $1/M$ for the first few correlation functions (the integral is not to
be performed numerically, of course, but using Wick's rule together with the covariance
(\ref{cov})). This leads to the results in (\ref{C1})-(\ref{C2}). Notice that letting $N\to 0$ in this context presents no difficulty.

The remainder of this paper is dedicated to the exact solution of the matrix integral
(\ref{ourintegral}), and the calculation of its limit as $N\to 0$.

\subsection{Exact Solution}
\subsubsection{Angular integration}

Introduce the singular value decomposition $Z=UDV$, where $D$ is real, positive and diagonal while
$U$ and $V$ are unitary. Let $X=D^2$ be a matrix with the same eigenvalues as $ZZ^\dag$,
and denote these eigenvalues by $x_i$, $1\leq i\leq N$. It is known \cite{morris} that the measure $dZ$
is expressed in these new variables as \be dZ=c_N|\Delta(X)|^2d\vec{x}dUdV,\ee where
$c_N$ depends only on the dimension, $dU$ is the normalized Haar measure on the unitary
group $\mathcal{U}(N)$, and the Vandermonde squared is the Jacobian of the
transformation. This is a generalization of the transformation from cartesian to polar
coordinates in the complex plane. We shall first perform the angular integration over $U$
and $V$.

A minor point to be mentioned is that $dV$ is not the same as the normalized Haar
measure. This is related to the fact that in the singular value decomposition there is a
certain ambiguity, as we may freely conjugate $D$ by a diagonal unitary matrix. The
matrix $V$ is thus uniquely determined only as an element of the coset
$\mathcal{U}(N)/[\mathcal{U}(1)]^N$. However, the functions we shall integrate,
polynomials in matrix elements as those in Section 3.3, are all invariant under
multiplication by a diagonal unitary matrix, and in this context $dV$ behaves just like the Haar measure, up to normalization.

The only part of the integral in (\ref{ourintegral}) that depends on the angular variables $U$ and $V$ is the last product.
Thus, the angular integral is \be \mathcal{A}=\int dUdV
\prod_{k=1}^n\sum_{j_k,m_k}U_{i_kj_k}D_{j_k}V_{j_ko_k}
V^{\dag}_{o_{\xi(k)}m_k}D_{m_k}U^\dag_{m_k i_k},\ee which can be expressed terms of
Weingarten functions as \be \mathcal{A}=\sum_{\sigma\tau\rho\theta\in S_{n}}{\rm
Wg}^{U}_N(\rho\theta^{-1}){\rm Wg}^{U}_N(\tau\sigma^{-1})p_{\tau^{-1}\theta}(X)
\delta_\sigma[i, i]\delta_\rho[o, \xi(o)],\ee where we have used that \be
\prod_{k=1}^n\sum_{j_k,m_k}D_{j_k}D_{m_k}\delta_\tau[j,m]\delta_\theta[j,m]=
\prod_{k=1}^n\sum_{j_k}x_{j_k}\delta_{\tau^{-1}\theta}[j,j]=p_{\tau^{-1}\theta}(X). \ee

The quantity we are after, Eq.(\ref{sumio}), requires summation over the indices
$\vec{i}$ and $\vec{o}$. It is easy to see that \be \sum_{i_1,\cdots,i_n=1}^{M}
\delta_\sigma[i, i]=p_{\sigma}(1^M), \quad \sum_{o_1,\cdots,o_n=1}^{M} \delta_\rho[o,
\xi(o)]=p_{\rho\xi}(1^M).\ee Notice that the channel labels in the original matrix
integral (\ref{ourintegral}) are all constrained to be between $1$ and $N$. Nevertheless,
we are summing them from $1$ to $M$. We are thus assuming $N\geq M$. However, this will
not deter us from letting $N\to 0$ later.

Once we expand \be p_{\tau^{-1}\theta}(X)=\sum_{\lambda\vdash
n}\chi_\lambda(\tau^{-1}\theta)s_\lambda(X),\ee we get \be
\sum_{\vec{i},\vec{o}}\mathcal{A}=\sum_{\lambda\vdash
n}\sum_{\sigma\tau\rho\theta\in S_{n}}{\rm
Wg}^{U}_N(\rho\theta^{-1}){\rm
Wg}^{U}_N(\tau\sigma^{-1})\chi_\lambda(\tau^{-1}\theta)s_\lambda(X)p_{\sigma}(1^M)p_{\rho\xi}(1^M).\ee Using identity
Eq.(\ref{convo}) twice leads to \be \sum_{\sigma\rho\in S_{n}}\sum_{\lambda\vdash
n}\frac{1}{([N]^\lambda)^2}\chi_\lambda(\rho^{-1}\sigma)p_{\sigma}(1^M)p_{\rho\xi}(1^M).\ee
Using another identity, \be \sum_{\sigma\in S_n}
\chi_\lambda(\rho^{-1}\sigma)p_\sigma(1^M)=\chi_\lambda(\rho)[M]^\lambda,\ee twice
finally leads to \be
 \sum_{\vec{i},\vec{o}}\mathcal{A}= \sum_{\lambda\vdash
n}\chi_\lambda(\xi)\left(\frac{[M]^\lambda}{[N]^\lambda}\right)^2s_\lambda(X).\ee

\subsubsection{Eigenvalue integration}

So far, the quantity we are after is given by \be\label{G}  \mathcal{G}_n=\sum_{\vec{i},\vec{o}}
G_n=\sum_{\lambda\vdash
n}\chi_\lambda(\xi)\left(\frac{[M]^\lambda}{[N]^\lambda}\right)^2\mathcal{R},\ee where
$\mathcal{R}$ is the radial integral over the eigenvalues of $ZZ^ \dag$. It is equal to
\be \mathcal{R}=\frac{c_N}{M\mathcal{Z}}\int_0^1 d\vec{x} \det{(1-X)^{M}}e^{Mi\epsilon{\rm
Tr}\left[\frac{ X}{1-X}\right]}|\Delta(x)|^2s_{\lambda}(X),\ee where we have used that
\be e^{-M\sum_{q\ge 1}\frac{(1-iq\epsilon)}{q}\tr
X^q}=\det{\left[(1-X)^{M}\right]}e^{Mi\epsilon\tr\left(\frac{ X}{1-X}\right)}.\ee From
the well known Schur function expansion, \be e^{Mi\epsilon{\rm
Tr}\left(\frac{X}{1-X}\right)}=\sum_{m=0}^\infty \frac{(Mi\epsilon)^m}{m!}\sum_{\mu\vdash
m}d_\mu s_\mu\left(\frac{X}{1-X}\right),\ee we get
\be\label{radial}\mathcal{R}=\sum_{m=0}^\infty \frac{(Mi\epsilon)^m}{m!}\sum_{\mu\vdash
m}d_\mu\mathcal{I}_{\lambda,\mu},\ee where \be
\mathcal{I}_{\lambda,\mu}=\frac{c_N}{\mathcal{Z}}\int_0^1 d\vec{x} \det
(1-X)^{M}|\Delta(x)|^2s_\mu\left(\frac{X}{1-X}\right)s_\lambda(X).\ee

Using the determinantal form of the Schur functions, the identity \be
\Delta\left(\frac{X}{1-X}\right)=\frac{\Delta(X)}{\det(1-X)^{N-1}}\ee and the integral
identity Eq.(\ref{andreief}), one can show that \be \mathcal{I}_{\lambda,\mu}=
\frac{c_NN!}{\mathcal{Z}}\det\left(\frac{(M-\mu_j+j-1)!(\lambda_i-i+\mu_j-j+2N)!}{(M+2N+\lambda_i-i)!}\right).\ee
Two factorials can be taken out of the determinant, and we can write \be
\mathcal{I}_{\lambda,\mu}=\frac{c_NN!}{\mathcal{Z}}\prod_{j=1}^N
\frac{(M-\mu_j+j-1)!}{(M+2N+\lambda_j-j)!}\det\left((\lambda_i-i+\mu_j-j+2N)!\right).\ee
Introducing $(M+j-1)!$ in the product, we get
\be\mathcal{I}_{\lambda,\mu}=\frac{c_NN!}{\mathcal{Z}}\frac{1}{[M]_\mu}\prod_{j=1}^N
\frac{(M+N-j)!}{(M+2N+\lambda_j-j)!}\det\left((\lambda_i-i+\mu_j-j+2N)!\right).\ee

\subsection{The $N\to 0$ limit}

We must now take the $N\to 0$ limit. This is a delicate procedure. We can only do it for
quantities that are analytic functions of $N$. For example, using the singular value
decomposition, the normalization constant (\ref{norm}) becomes \be \mathcal{Z}=c_N\int_0^\infty
dxe^{-M(1-i\epsilon)\tr
X}|\Delta(x)|^2=\frac{c_N}{[M(1-i\epsilon)]^{N^2}}\prod_{j=1}^Nj!(N-j)!.\ee It is
perfectly fine to take the limit in the denominator. In the rest of the expression, we
must leave $N$ intact for now. In this sense, we write \be\label{norm2} \mathcal{Z}\to
c_N\prod_{j=1}^Nj!(N-j)!.\ee

The quantity $\mathcal{I}_{\lambda,\mu}$ contains the factor \be \prod_{j=1}^N
\frac{(M+N-j)!}{(M+2N+\lambda_j-j)!}.\ee First, we let $N\to 0$ inside the product, to
get\be \prod_{j=1}^N \frac{(M-j)!}{(M+\lambda_j-j)!}.\ee This still depends on $N$ via
the limit of the product. However, $\lambda_j=0$ for $j>\ell(\lambda)$. Hence, if we
assume $N\geq\ell(\lambda)$, we can write this as \be \prod_{j=1}^{\ell(\lambda)}
\frac{(M-j)!}{(M+\lambda_j-j)!}=\frac{1}{[M]^\lambda},\ee which is independent of $N$.
Now, in all rigor we are not allowed to take $N\to 0$ after assuming
$N\geq\ell(\lambda)$. We do it anyway, and write \be\mathcal{I}_{\lambda,\mu}\to
\frac{N!}{\mathcal{Z}}\frac{1}{[M]_\mu[M]^\lambda}\det\left((\lambda_i-i+\mu_j-j+2N)!\right).\ee

Further, we factor out the smallest factor from each row of the determinant, producing
$\prod_{j=1}^N(N+\lambda_j-j+\mu_N)!$. If we assume that $N>\ell(\mu)$, then $\mu_N=0$.
Hence, using (\ref{norm2}), \be\mathcal{I}_{\lambda,\mu}\to
\frac{N!}{[M]_\mu[M]^\lambda}\prod_{j=1}^N\frac{(N+\lambda_j-j)!}{(N-j)!j!}
\det\left(\frac{(\lambda_i-i+\mu_j-j+2N)!}{(\lambda_i-i+N)!}\right).\ee We again consider
$N\geq\ell(\lambda)$ first and $N\to 0$ later, to arrive
at\be\label{II}\mathcal{I}_{\lambda,\mu}\to
\frac{[N]^\lambda}{[M]_\mu[M]^\lambda}\frac{1}{\prod_{j=1}^{N-1}j!}\det\left(\frac{(\lambda_i-i+\mu_j-j+2N)!}{(\lambda_i-i+N)!}\right).\ee

\subsubsection{The determinant}

We need to consider the determinant
\be\mathcal{D}=\det\left(\frac{(\lambda_i-i+\mu_j-j+2N)!}{(\lambda_i-i+N)!}\right)
=\det\left(\frac{(a_i+b_j)!}{a_i!}\right),\ee where \be a_i=\lambda_i-i+N, \quad
b_j=\mu_j-j+N.\ee Each column consists of raising factorials, i.e. we have \be
(a_i+1)(a_i+2)\cdots(a_i+b_j)=\frac{[a_i]^{b_j+1}}{a_i}.\ee We therefore expand each
column using identity Eq.(\ref{Stir}), in terms of unsigned Stirling numbers of the first
kind,
\be \frac{[a_i]^{b_j+1}}{a_i}=\sum_{k_j=1}^{b_j+1}\left[\begin{array}{c}b_j+1 \\
k_j \end{array}\right]a_i^{k_j-1}=\sum_{k_j=0}^{b_j}\left[\begin{array}{c}b_j+1 \\
k_j+1 \end{array}\right]a_i^{k_j}.\ee The determinant is then given by
\be\mathcal{D}=\prod_{j=1}^N\sum_{k_j=0}^{b_j}\left[\begin{array}{c}b_j+1
\\k_j+1 \end{array}\right] \det\left(a_i^{k_j}\right).\ee Introducing $k_j=\omega_j-j+N$
we have \be\mathcal{D}=\prod_{j=1}^N\sum_{\omega_j=j-N}^{\mu_j}
\left[\begin{array}{c}\mu_j-j+N+1 \\ \omega_j-j+N+1 \end{array}\right]
\det\left(a_i^{\omega_j-j+N}\right).\ee

Notice that $\omega$ is not a partition, since its elements are not necessarily ordered,
and they can be negative. Still, the last determinant, if it does not vanish, can be
turned into a Schur function by simply re-ordering the columns. Let $\widetilde{\omega}$
be the partition that is created in this way, and $|\widetilde{\omega}|$ the number it partitions. For instance, if $\omega=(1,1,-1,1)$ we
have \be \det\left(\begin{matrix} a_i^N & a_i^{N-1} & a_i^{N-4} & a_i^{N-3} & a_i^{N-5}
\cdots\end{matrix}\right) =-\det\left(\begin{matrix} a_i^N & a_i^{N-1} & a_i^{N-3}
\cdots\end{matrix}\right),\ee so the corresponding partition is
$\widetilde{\omega}=(1,1)$ and $|\widetilde{\omega}|=2$. As we can see, the reordering of the columns may lead to a
change in sign. Let $\eta(\omega)$ denote this sign, so that \be
\det\left(a_i^{\omega_j-j+N}\right)=\eta(\omega)\Delta(a)s_{\widetilde{\omega}}(a)=\eta(\omega)\frac{d_\lambda}{n!}[N]^\lambda
s_{\widetilde{\omega}}(a)\prod_{j=1}^{N-1}j!.\ee

We must consider the $N\to 0$ limit of \be s_{\widetilde{\omega}}(a)=\frac{1}{|\widetilde{\omega}|!}\sum_{\rho\vdash
|\widetilde{\omega}|}|\mathcal{C}_\rho|\chi_{\widetilde{\omega}}(\rho)p_\rho(a).\ee
 The
limit of $p_\rho(a)$ can be obtained simply removing from this quantity everything that
scales with $N$: \be\lim_{N\to
0}p_\rho(\{\lambda_i-i+N\})=\prod_{q=1}^{\ell(\rho)}\left(\sum_{i=1}^{\ell(\lambda)}(\lambda_i-i)^q-(-i)^q\right)\equiv
f_\rho(\lambda).\ee We can finally write \be
\frac{\mathcal{D}}{\prod_{j=1}^{N-1}j!}\to\frac{d_\lambda}{n!}[N]^\lambda
F_{\lambda,\mu},\ee where the function $F_{\lambda,\mu}$ is given by \be
F_{\lambda,\mu}=\prod_{j=1}^{\ell(\mu)}\sum_{\omega_j=j-\ell(\mu)}^{\mu_j}
\left[\begin{array}{c}\mu_j-j+1 \\ \omega_j-j+1 \end{array}\right]
\frac{\eta(\omega)}{|\widetilde{\omega}|!}\sum_{\rho\vdash
|\widetilde{\omega}|}|\mathcal{C}_\rho|\chi_{\widetilde{\omega}}(\rho)f_\rho(\lambda).\ee

\subsubsection{Final Result}

It is time to put the pieces back together. We have to plug the limiting value of
$\mathcal{D}$ into the expression for $\mathcal{I}_{\lambda,\mu}$, Eq.(\ref{II}), put
this into the expression for the radial integral, Eq. (\ref{radial}), and finally arrive
at the quantity we want, which is $\mathcal{G}_n$, Eq.(\ref{G}). After some cancelations,
we get that the limit as $N\to 0$ of $\mathcal{G}_n$, which is nothing but the
semiclassical expression for the correlation function $C_n(\epsilon)$, is given by \be
\lim_{N\to 0}\mathcal{G}_n=C_n(\epsilon)=\frac{1}{Mn!}\sum_{m=0}^\infty
\frac{(Mi\epsilon)^m}{m!}\sum_{\mu\vdash m}\sum_{\lambda\vdash n}d_\lambda
d_\mu\chi_\lambda(\xi)\frac{[M]^\lambda}{[M]_\mu} F_{\lambda,\mu}.\ee

This expression is perhaps not as simple we one might hope for, specially the
$F_{\lambda,\mu}$ part. This complication is probably due to the
fact that we are using a Taylor series in $\epsilon$. We know that, at each order in
$1/M$, the correlation functions are rational functions of $\epsilon$, with the
denominator being a power of $(1-i\epsilon)$. Maybe if this fact could be explicitly
incorporated into the calculation somehow, the resulting expression would be more
manageable.

For the simplest correlation function, explicit calculations suggest that the following
expression holds: \be C_1(\epsilon)=\sum_{n=1}^\infty
\frac{(Mi\epsilon)^n}{n}\sum_{k=0}^{n-1}\frac{1}{[M+k]_n},\ee which is indeed in
agreement with the first $3$ orders in $1/M$ as computed from (\ref{C1}).

The average value of linear moments $\mathcal{M}_m$ is given by \be \langle \mathcal{M}_m\rangle=\frac{\tau_D^m}{i^mm!}\left[\frac{d^m}{d\epsilon^m}\sum_{n=1}^m
(-1)^{m-n}{m \choose n}C_n(\epsilon)\right]_{\epsilon=0}.\ee Therefore, if the identity  \be\label{identity} \frac{1}{n!}\sum_{n=1}^m
(-1)^{m-n}{m \choose n} \sum_{\lambda\vdash n}d_\lambda\chi_\lambda(n)[M]^\lambda F_{\lambda,\mu}=[M]^\mu\chi_\mu(m),\ee is true,
then the semiclassical formula for $\langle \mathcal{M}_m\rangle$
becomes exactly equal to the RMT prediction (\ref{Mn}). We have checked that
(\ref{identity}) indeed holds for all $\mu\vdash m$ up to $m=8$ (in doing so one needs only deal with hook partitions, for otherwise the character $\chi_\lambda(n)$ vanishes). This guarantees agreement between the semiclassical and RMT calculations up to the first 8 moments. Incidentally, since both expressions for $\mathcal{M}_m$ are written as a sum over $\langle s_\lambda\rangle$, this suggests that the agreement between these approaches extends to all Schur functions, and hence to all statistics, as would be expected.

\section*{Acknowledgments}

Financial support from Conselho Nacional de Desenvolvimento Cient\'ifico e Tecnol\'ogico (CNPq) is gratefully acknowledged.


\begin{thebibliography}{99}

\bibitem{wigner} E.P. Wigner, Phys. Rev. {\bf 98}, 145 (1955).
\bibitem{smith} F.T. Smith, Phys. Rev. {\bf 118}, 349 (1960).

\bibitem{review2} C.A.A. de Carvalho and H.M. Nussenzveig, Phys. Rep. {\bf 364}, 83 (2002).

\bibitem{cavities1} C. Dembowski, B. Dietz, T. Friedrich, H.-D. Gr\"af, A. Heine,
C. Mej\'ia-Monasterio, M. Miski-Oglu, A. Richter and T. H. Seligman. Phys. Rev. Lett. {\bf 93}, 134102 (2004).

\bibitem{cavities2} B. Dietz, T. Friedrich, H. L. Harney, M. Miski-Oglu, A. Richter, F. Sch\"afer,
and H. A. Weidenm\"uller, Phys. Rev. Lett. {\bf 98}, 074103 (2007).

\bibitem{cavities3}  A. Backer, R. Ketzmerick, S. L\"ock, M. Robnik, G. Vidmar, R. H\"ohmann,
U. Kuhl, and H.-J. St\"ockmann, Phys. Rev. Lett. {\bf 100}, 174103 (2008).

\bibitem{cavities4} S. Hemmady, J. Hart, X, Zheng, T. M. Antonsen Jr., E. Ott and S. M.
Anlage, Phys. Rev. B {\bf 74}, 195326 (2006).

\bibitem{dots1} Y. Alhassid, Rev. Mod. Phys. {\bf 72}, 895 (2000).

\bibitem{dots2} S. Oberholzer, E.V. Sukhorukov e C. Schonenberger, Nature {\bf 415}, 765 (2002).

\bibitem{dots3} J. Bylander, T. Duty e P. Delsing, Nature {\bf 434}, 361 (2005);

\bibitem{dots4} S. Gustavsson, R. Leturcq, B. Simovi\v{c}, R. Schleser, T. Ihn, P. Studerus, K. Ensslin, D.
C. Driscoll, and A. C. Gossard, Phys. Rev. Lett. {\bf 96}, 076605 (2006).

\bibitem{nuclear} G.E. Mitchell, A. Richter and H.A. Weidenm\"uller, Rev. Mod. Phys. {\bf 82}, 2845 (2010).

\bibitem{caio1} C. H. Lewenkopf and H. A. Weidenm\"uller, Ann. Phys. {\bf 212}, 53 (1991).

\bibitem{caio2} C.H. Lewenkopf and R. O. Vallejos, Phys. Rev. E {\bf 70}, 036214 (2004).

\bibitem{beenakker} C.W.J. Beenakker, Rev. Mod. Phys. {\bf 69}, 731 (1997).

\bibitem{prb78mn2008} M. Novaes, Phys. Rev. B {\bf 78}, 035337 (2008).

\bibitem{prb80bak2009} B.A. Khoruzhenko, D.V. Savin, and H.J. Sommers, Phys. Rev. B {\bf 80}, 125301 (2009).

\bibitem{macedo} F.A.G. Almeida, S. Rodr\'iguez-P\'erez and A.M.S. Mac\^edo, Phys. Rev. B {\bf 80}, 125320 (2009).

\bibitem{sigma1} J.J.M. Verbaarschot, H.A. Weidenm\"uller and M.R. Zirnbauer, Phys. Rep. {\bf 129}, 367 (1985).

\bibitem{sigma2} N. Lehmann, D. V. Savin, V. V. Sokolov and H.-J. Sommers, Physica D {\bf 86}, 572 (1995).

\bibitem{sigma3} Y. V. Fyodorov and H.-J. Sommers, J. Math. Phys. {\bf 38}, 1918 (1997).

\bibitem{sigma4} S. Kumar, A. Nock, H.-J. Sommers, T. Guhr, B. Dietz, M. Miski-Oglu, A. Richter, and F.
Sch\"afer, Phys. Rev. Lett. {\bf 111}, 030403 (2013).

\bibitem{brouwer} P.W. Brouwer, K.M. Frahm and C.W.J. Beenakker, Phys. Rev. Lett. {\bf 78}, 4737
(1997).

\bibitem{brouwer2} P.W. Brouwer, K.M. Frahm and C.W.J. Beenakker, Waves Rand. Media {\bf 9}, 91 (1999).

\bibitem{Mis2} D.V. Savin, Y.V. Fyodorov and H.-J. Sommers, Phys. Rev. E {\bf 63}, 035202 (2001).

\bibitem{majumdar} C. Texier and S.N. Majumdar, Phys. Rev. Lett. {\bf 110}, 250602 (2013).

\bibitem{simm1} F. Mezzadri and N. Simm, J. Math. Phys. {\bf 52}, 103511 (2011).
\bibitem{simm2} F. Mezzadri and N. Simm, J. Math. Phys. {\bf 53}, 053504 (2012).
\bibitem{simm3} F. Mezzadri and N. Simm, Comm. Math. Phys. {\bf 324}, 465 (2013).
\bibitem{garcia} A.M. Mart\'inez-Arg\"uello, M. Mart\'inez-Mares and J.C. Garc\'ia, J. Math. Phys. {\bf 55}, 081901 (2014).

\bibitem{dima} Y.V. Fyodorov and D. V. Savin, Chapter 34 in The Oxford Handbook of Random Matrix Theory
(Oxford, 2011), G. Akemann, J. Baik and P. Di Francesco (Editors).

\bibitem{jalabert} R.A. Jalabert, H.U. Baranger and A.D. Stone, Phys. Rev. Lett. {\bf 65}, 2442 (1990).

\bibitem{berry} M.V. Berry, Proc. Roy. Soc. A {\bf 400}, 229 (1985).
\bibitem{cvitanovic} P. Cvitanovi\'c and B. Eckhardt, J. Phys. A {\bf 24}, L237 (1991).
\bibitem{hannay} J. H. Hannay and A. M. Ozorio de Almeida, J. Phys. A {\bf 17}, 3429 (1984).

\bibitem{sieber} K. Richter and M. Sieber, Phys. Rev. Lett. {\bf 89}, 206801 (2002).

\bibitem{muller} S. M\"uller, S. Heusler, P. Braun and F. Haake, New J. Phys. {\bf 9}, 12 (2007).

\bibitem{greg} G. Berkolaiko and J. Kuipers, Phys. Rev. E {\bf 85}, 045201 (2012).

\bibitem{novaes1} M. Novaes, Europhys. Lett. {\bf 98}, 20006 (2012).

\bibitem{balian} R. Balian and C. Bloch, Ann. Phys. {\bf 85}, 514 (1974).

\bibitem{altmann} E.G. Altmann, J.S.E. Portela and T. T\'el, Rev. Mod. Phys. {\bf 85}, 869
(2013).

\bibitem{eckhardt} B. Eckhardt, Chaos {\bf 3}, 613 (1993).

\bibitem{vallejos} R.O. Vallejos, A.M. Ozorio de Almeida and C.H. Lewenkopf, J. Phys. A {\bf 31}, 4885 (1998).

\bibitem{KS1} J. Kuipers and M. Sieber, Nonlinearity {\bf 20}, 909 (2007).

\bibitem{KS2} J. Kuipers and M. Sieber, Phys. Rev. E {\bf 77}, 046219 (2008).

\bibitem{caio3} C.H. Lewenkopf and R.O. Vallejos, J. Phys. A {\bf 37}, 131 (2004).

\bibitem{berko1} G. Berkolaiko and J. Kuipers, J. Phys. A {\bf 43}, 035101 (2010).

\bibitem{berko2} G. Berkolaiko and J. Kuipers, New J. Phys {\bf 13}, 063020 (2011).

\bibitem{new} Personal communication.

\bibitem{novaes2} M. Novaes, J. Phys. A {\bf 46}, 502002 (2013).

\bibitem{andreev1} J. Kuipers, D. Waltner, C. Petitjean, G. Berkolaiko and K. Richter, Phys. Rev. Lett. {\bf 104}, 027001 (2010).

\bibitem{andreev2} J. Kuipers, T. Engl, G. Berkolaiko, C. Petitjean, D. Waltner and K. Richter, Phys. Rev. B {\bf 83}, 195315 (2011).

\bibitem{ehren1} I. Adagideli, Phys. Rev. B {\bf 68} 233308 (2003).
\bibitem{ehren2} S. Rahav and P.W. Brouwer, Phys. Rev. Lett. {\bf 95}, 056806 (2005).
\bibitem{ehren3} R.S. Whitney and Ph. Jacquod, Phys. Rev. Lett. {\bf 96}, 206804 (2006).
\bibitem{ehren4} D. Waltner, J. Kuipers and K. Richter, Phys. Rev. B {\bf 83}, 195315 (2011).

\bibitem{samuel} S. Samuel, J. Math. Phys. {\bf 21}, 2695 (1980).
\bibitem{esposti} M. Degli Esposti and A. Knauf, J. Math. Phys. {\bf 45}, 4957 (2004).
\bibitem{collins} B. Collins, Int. Math. Res. Not. {\bf 17}, 953 (2003).

\bibitem{R2c} S. M\"uller, S. Heusler, P. Braun, F. Haake, and A. Altland, Phys. Rev. E {\bf 72}, 046207 (2005).

\bibitem{GregJack1} G. Berkolaiko and J. Kuipers, J. Math. Phys. {\bf 54}, 112103 (2013)
\bibitem{GregJack2} G. Berkolaiko and J. Kuipers, J. Math. Phys. {\bf 54}, 123505 (2013).

\bibitem{difrancesco} P. Di Francesco, in Applications of Random Matrices in Physics
(Springer, 2006), \'E. Brezin and V. Kazakov (Editors).

\bibitem{bouttier} J. Bouttier, Chapter 26 in The Oxford Handbook of Random Matrix Theory
(Oxford, 2011), G. Akemann, J. Baik and P. Di Francesco (Editors).

\bibitem{morris} T.R. Morris, Nucl. Phys. B {\bf 356}, 703 (1991).

\end{thebibliography}
\end{document}